\documentclass[twocolumn]{aastex631}

\received{May 10}
\revised{July 1}
\accepted{July 10}

\usepackage{CJK}

\begin{document}

\title{5-25 $\mu$m Galaxy Number Counts from Deep JWST Data}

\shorttitle{Deep MIRI Number Counts}

\shortauthors{M. Stone et al.}

\correspondingauthor{Meredith A. Stone}
\email{meredithstone@arizona.edu}

\author[0000-0002-9720-3255]{Meredith A. Stone}
\affiliation{Steward Observatory, University of Arizona,
933 North Cherry Avenue, Tucson, AZ 85719, USA}

\author[0000-0002-8909-8782]{Stacey Alberts}
\affiliation{Steward Observatory, University of Arizona,
933 North Cherry Avenue, Tucson, AZ 85719, USA}

\author[0000-0003-2303-6519]{George H. Rieke}
\affiliation{Steward Observatory, University of Arizona,
933 North Cherry Avenue, Tucson, AZ 85719, USA}

\author[0000-0002-8651-9879]{Andrew J. Bunker}
\affiliation{Department of Physics, University of Oxford,
Denys Wilkinson Building, Keble Road, Oxford OX1 3RH, UK}

\author[0000-0002-6221-1829]{Jianwei Lyu (\begin{CJK}{UTF8}{gbsn}吕建伟\end{CJK})}
\affiliation{Steward Observatory, University of Arizona,
933 North Cherry Avenue, Tucson, AZ 85719, USA}

\author[0000-0003-4528-5639]{Pablo G. P\'erez-Gonz\'alez}
\affiliation{Centro de Astrobiolog\'ia (CAB), CSIC-INTA, Ctra. de Ajalvir km 4, Torrej\'on de Ardoz, E-28850, Madrid, Spain}

\author[0000-0003-4702-7561]{Irene Shivaei} \affiliation{Centro de Astrobiolog\'ia (CAB), CSIC-INTA, Ctra. de Ajalvir km 4, Torrej\'on de Ardoz, E-28850, Madrid, Spain}

\author[0000-0003-3307-7525]{Yongda Zhu}
\affiliation{Steward Observatory, University of Arizona,
933 North Cherry Avenue, Tucson, AZ 85719, USA}

\begin{abstract}

Galaxy number counts probe the evolution of galaxies over cosmic time, and serve as a valuable comparison point to theoretical models of galaxy formation. We present new galaxy number counts in eight photometric bands between 5 and 25 $\mu$m from the Systematic Mid-infrared Instrument Legacy Extragalactic Survey (SMILES) and the JWST Advanced Deep Extragalactic Survey (JADES) deep MIRI parallel, extending to unprecedented depth. By combining our new MIRI counts with existing data from Spitzer and AKARI, we achieve counts across 3-5 orders of magnitude in flux in all MIRI bands. Our counts diverge from predictions from recent semi-analytical models of galaxy formation, likely owing to their treatment of mid-infrared aromatic features. Finally, we integrate our combined JWST-Spitzer counts at 8 and 24 $\mu$m to measure the cosmic infrared background (CIB) light at these wavelengths; our measured CIB fluxes are consistent with those from previous mid-infrared surveys, but larger than predicted by some models based on TeV blazar data. 

\end{abstract}

\section{Introduction} \label{sec:intro}

The distribution of the number of sources detected in a given filter as a function of flux, known as source counts or number counts, probes galaxy evolution on the largest scales and provides a direct observational comparison to the results of cosmological simulations and semi-analytical models of galaxy formation and evolution. Unlike luminosity or mass functions, which can only be constructed with robust redshift constraints (difficult to obtain for huge samples of galaxies, and with their own uncertainties), deriving number counts requires only sufficient depth, area, and robust measurements of source fluxes and is therefore subject to fewer sources of calibration bias when comparing to simulations. Number counts can also be integrated to provide an estimate of the Extragalactic Background Light (EBL) at a given wavelength, and compared to estimates from modeling the interaction of high-energy gamma rays with lower-energy EBL photons.

Number counts at different wavelengths in the mid and far-infrared (IR) in particular have revealed a great deal about the evolution of galaxies across cosmic time. The IRAS and ISO missions revealed that the distribution of mid- and far-IR number counts differs significantly from the optical and near-IR. While optical and near-IR counts display a ``Euclidean" $dN/dS_\nu \propto S_\nu^{-2.5}$ slope down to faint fluxes, consistent with a nonevolving population of spiral and elliptical galaxies, counts at longer mid-infrared wavelengths display a characteristic ``bump" between fluxes of approximately $0.1$ and $0.5$ mJy before decaying at fainter fluxes, attributed to the rapidly evolving population of galaxies between the local Universe and $z\sim2$ \citep[see e.g.][]{Fazio2004}. In the 21st century, mid- and far-IR number counts have been probed to fainter and fainter limits, and the contributions of different galaxy populations inferred, with observations from later generations of IR space telescopes including Spitzer's IRAC \citep{Fazio2004, Treister2006, Barmby2008, Ashby2009, Ashby2015, Papovich2016, Lacy2021}, IRS \citep{Teplitz2011}, and MIPS \citep{Marleau2004, Papovich2004, Chary2004, Treister2006, RoccaVolmerange2007, Bethermin2010, Clements2011} instruments and the AKARI mission \citep{Pearson2010, Pearson2014, Murata2014, Davidge2017}. Prior to 2022, these missions had pushed the $80\%$ completeness limit to $<100$ $\mu$Jy at $8$ and $24$ $\mu$m with Spitzer, and somewhat higher in intermediate bands with AKARI.

With the launch of the James Webb Space Telescope (JWST), a wealth of new data will become available to probe number counts in the mid-infrared with MIRI. As number counts from MIRI data are brand-new and able to reach unprecedented depth and access sources fainter than ever observed, they provide critical constraints on the faint end of the distribution, a region that currently displays the most variation between different cosmological simulations and semi-analytical models \citep[see e.g.][]{Lacy2021}. These theoretical models are not yet calibrated to match JWST observations, so JWST data provides an independent test of model validity.

Already, mid-infrared number counts from MIRI have been published in the Stephan's Quintet ERS field \citep{Ling2022} at 7.7, 10, and 15 $\mu$m and in a portion of the Cosmic Evolution Early Release Science (CEERS) field in the Extended Groth Strip \citep{Wu2023, Kirkpatrick2023}. These data extend down to $80$\% completeness limits ranging from 0.25 $\mu$Jy at 7.7 $\mu$m to 13 $\mu$Jy at 21 $\mu$m, in some filters extending more than two orders of magnitude deeper than the deepest pre-JWST data available. However, the small area of these surveys (of order 5 arcmin$^2$ in each field) limited their ability to connect to previous observations at the bright end and leave them vulnerable to cosmic variance.

In this work, we present number counts derived from the Systematic Mid-infrared Instrument Extragalactic Survey (SMILES) in eight MIRI bands, supplemented by ultra-deep MIRI parallel imaging at 7.7$\,\mu$m observed as part of the JWST Advanced Deep Extragalactic Survey (JADES). The wider area ($\sim34$ arcmin$^2$) and complete mid-infrared wavelength coverage of SMILES complement the incredibly deep data of the MIRI JADES parallel and previously existing Spitzer data, allowing us to probe infrared emission across more than five orders of magnitude in flux and suppress cosmic variance by linking MIRI counts to those from wider Spitzer surveys. We introduce our observations in Section \ref{sec:data}, outlining our data reduction and completeness correction processes and photometric accuracy tests performed. We describe the construction of the number counts and present our results in Section \ref{sec:results}. In Section \ref{sec:discussion}, we compare to a number of state-of-the-art theoretical models of galaxy formation and evolution. We also integrate our 7.7 and 21 $\mu$m counts to constrain the cosmic infrared background light and compare the result with values deduced from TeV gamma ray observations. We summarize in Section \ref{sec:conclusions}.

\section{Data}\label{sec:data}

The primary datasets for this work are MIRI imaging in the GOODS-S field from the Systematic Mid-infrared Instrument Legacy Extragalactic Survey (SMILES, PID 1207) at 5.6 - 25.5 $\mu$m \citep{Rieke2024, Alberts2024} and the ultra-deep 7.7 $\mu$m parallel (S. Alberts, et al., in prep) from the JWST Advanced Deep Extragalactic Survey \citep[JADES, PID 1180;][]{Eisenstein2023}.  SMILES is comprised of 15 pointings ($\sim34$ arcmin$^2$) of moderate depth exposures (10.7 - 36.4 min) in 8 MIRI filters (excluding only F1130W), reaching $5\sigma$ point source sensitivities of 0.20, 0.19, 0.38, 0.59, 0.68, 1.7, 2.8 and 16 $\mu$Jy for F560W, F770W, F1000W, F1280W, F1500W, F1800W, F2100W, and F2550W respectively, measured in apertures containing 65\% of the encircled energy for each PSF and then aperture corrected (see Section~\ref{sec:phot}).  

The JADES deep MIRI parallel is four pointings ($\sim10$ arcmin$^2$) totalling 43 hours of science time per pointing, reaching a $5\sigma$ point source sensitivity of 21 nJy in F770W.  The $5\sigma$ point source sensitivities are shown in Table~\ref{tab:complete} and are generally 2-2.5x lower than predicted by ETC v3.0 up to F1800W.

The majority of the area of these two surveys fall on deep and medium JADES NIRCam imaging \citep{Eisenstein2023} as well as the rich ancillary data in GOODS-S from  facilities such as the Hubble Space Telescope \citep[HST, e.g.][]{Giavalisco2004, Grogin2011}.

\subsection{Data Reduction}\label{sec:datareduction}

The data reduction for SMILES and the JADES MIRI parallel are broadly similar; a full description can be found in \cite{Alberts2024}
Both datasets were reduced with v.1.12.5 of the JWST calibration pipeline \citep{Bushouse2023} using JWST Calibration Reference System (CRDS) context jwst\_1188.pmap.

We supplement the standard pipeline with custom, external routines to remove warm pixels missed by the pipeline, ``super'' background subtraction \citep{Perez2024}, and apply astrometry corrections, matched to GAIA through the JADES NIRCam catalog \citep{Rieke2023b}. The final astrometry is accurate to $0.1-0.2\arcsec$  ($1\sigma$, per axis) for F560W-F2100W and $0.4\arcsec$ for F2550W, as imaging in this filter is relatively shallow compared to the rest of the survey, with fewer detections suitable for deriving an astrometric correction \citep{Alberts2024}. 

\subsection{Source Extraction and Photometry}\label{sec:phot}

Blind MIRI photometric catalogs are created with a modified version of the JADES photometric pipeline \citep[][Robertson et al., in prep]{Rieke2023b}.  Object detection is done using a stacked F560W and F770W SNR image for SMILES and F770W only for the JADES parallel; these detection images are then used to create a robust segmentation map through an iterative process that starts at a low threshold for detection and then performs deblending, cleaning to remove spurious noise spikes, and faint source detection.

Circular aperture photometry (in apertures of diameter $0.5, 0.6, 0.7$, and $1.0\arcsec$) is then measured from the final, clean segmentation map using {\sc photutils} with the windowed centroiding algorithm from Source Extractor.  Kron photometry (2.5x scaled) is also measured using a stacked signal map to define source centroids.  Kron apertures smaller than our smallest circular aperture ($d=0.5\arcsec$) are replaced with that circular aperture photometry. SMILES photometry in filters longer than F770W is measured using the apertures defined by the stacked F560W+F770W detection image.  We assess whether this approach misses long wavelength-only detections and/or changes the derived long wavelength flux densities; we find that these effects are small and that our photometric catalog is robust \citep{Alberts2024}.

Aperture corrections are applied to both the circular aperture and Kron aperture fluxes using custom point spread functions (PSFs). We derive the aperture corrections for the latter by measuring the fraction of the PSF flux which falls outside each Kron aperture, to account for the size and shape of the PSF. For F1000W-F2550W, PSFs are created using WebbPSF \citep{Perrin2014}.  For F560W and F770W, the PSF modeling is complicated by the cross-like imaging artifact known as the ``cruciform" \citep{Gaspar2021} and so empirical PSFs are constructed using high dynamic range imaging of stars obtained during JWST commissioning (A. G\'asp\'ar, private communication).  Finally, photometric uncertainties are obtained using randomly placed apertures across the mosaics, accounting for local exposure time and correlated pixel noise \citep[e.g.][]{Whitaker2011, Rieke2023b}.

The final SMILES photometric catalog has 3,591 sources with SNR$\,>4$ in either the F560W or F770W filter.  The JADES MIRI parallel catalog has 2,860 sources with SNR$\,>4$ in F770W.

\subsection{Completeness and Photometric Accuracy}\label{sec:completeness}

\begin{figure}
    \centering
    \includegraphics[width=\columnwidth]{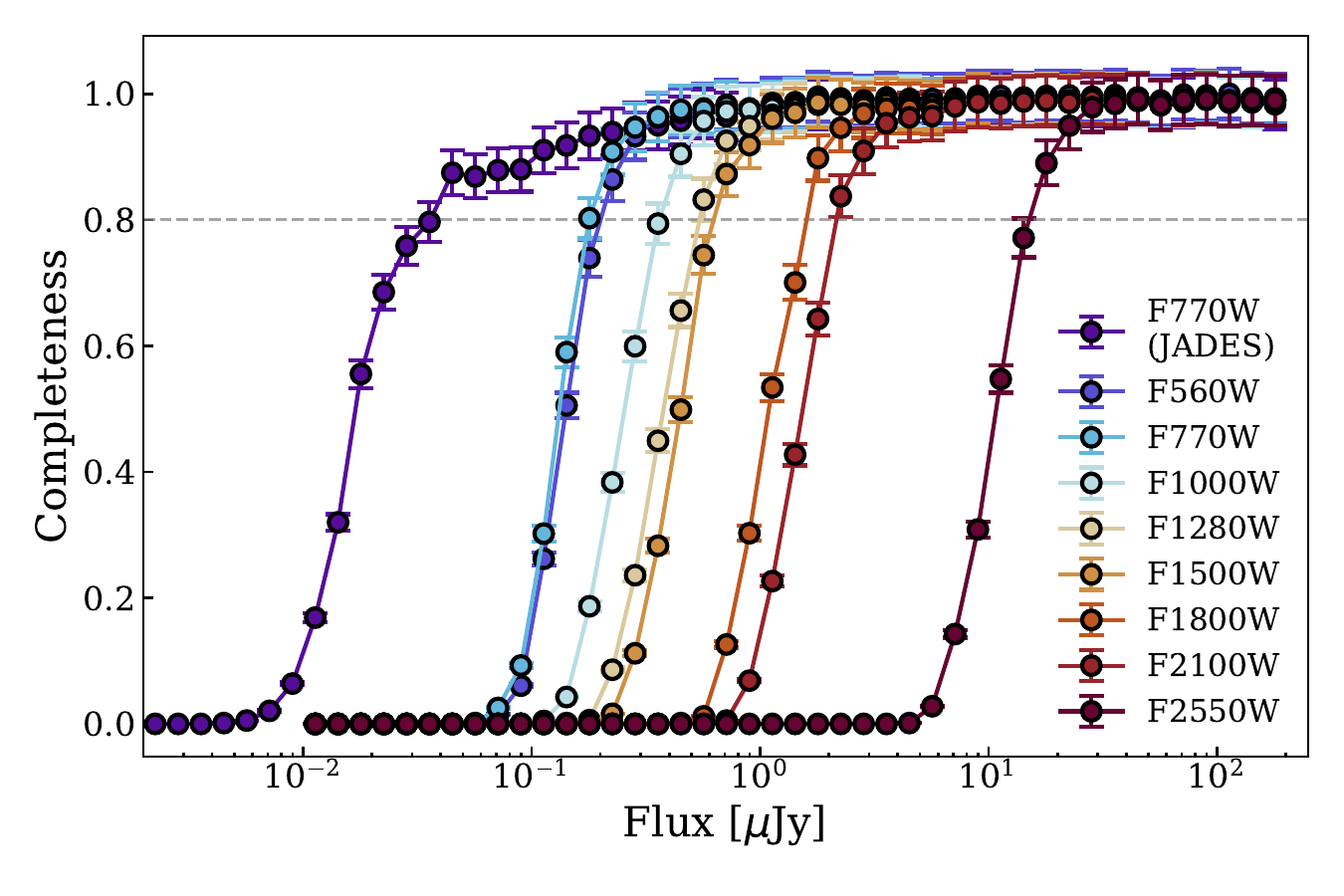}
    \caption{Completeness fraction as a function of flux for our MIRI data, including the JADES parallels F770W data (purple, leftmost points) and  the eight bands observed in the SMILES survey. Error bars are Poissonian. The 80\% completeness threshold, which is the lower limit to which we report our number counts, is shown as a gray dashed line.}
    \label{fig:complete}
\end{figure}

We measure the completeness $-$ the probability of measuring a source at a given flux density given the mosaic noise properties $-$ in our images by inserting and measuring the recovery rate of artificial point sources constructed from our PSFs (Section~\ref{sec:phot}). We derive the completeness using point sources because faint galaxies are predominantly intrinsically small and un- or marginally-resolved by MIRI, while MIRI-resolved sources tend to be bright. Bright extended sources are in a flux regime where the completeness corrections are negligible (see Figure \ref{fig:complete}), and we do not expect faint, extended sources (for which our completeness will be overestimated) to be numerous enough to significantly affect our counts.

For SMILES, artificial sources are inserted in 43 flux bins with a width of 0.1 dex ranging from 0.01-200 $\mu$Jy.  Within each bin, the flux of each artificial source is chosen at random.  For each filter and flux bin, we generate 10 mosaics with 150 artificial sources ($<5\%$ of the real source catalog) inserted into random positions (with masking of the map edges), totaling 1,500 artificial sources per filter per flux bin: a number that ensures robust Gaussian statistics while limiting computational time requirements.  Photometric catalogs are then created for each altered mosaic following the procedure outlined in Section~\ref{sec:phot}.  Artificial sources are considered recovered if they are found within $0.2\arcsec$ of the inserted position and the measured flux density is detected at $>2\sigma$ and within $50\%$ of the input flux. The latter accounts for flux boosting of faint sources from brighter neighbors under the approximation that galaxies are randomly distributed.  As galaxies are in reality clustered, this is an underestimation, but we do not expect this to have a significant impact on our measured number counts given the small MIRI beam \citep{Viero2013}.

This procedure is repeated for the JADES MIRI parallel with 53 flux bins ranging from 0.001 - 200$\mu$Jy with 0.1 dex binwidth.  Given the smaller area of the parallel, we inject 50 artificial sources at a time into 30 mosaics, again producing 1,500 artificial sources per flux bin.  The resulting completeness  curves for both surveys are shown in Figure~\ref{fig:complete} and the $80\%$ completeness level is reported in Table~\ref{tab:complete}. The uncertainties on the completeness curves are the Poissonian, $\sqrt{N}$ error. A completeness of $80\%$ in each band is reached at approximately the $5\sigma$ detection limit. This high completeness at faint fluxes is attributable to our careful background subtraction (Section~\ref{sec:datareduction}).  For the rest of this work, only sources above the $80\%$ completeness limit will be considered.

\begin{deluxetable}{lccc}
\tablecaption{80\% completeness limits from SMILES and JADES in the MIRI bands. \label{tab:complete}}
\tablewidth{0pt}
\tablehead{
\colhead{Band} & \colhead{$5\sigma$ Sensitivity} & \colhead{80\% Completeness} & \colhead{Area}\\[-0.25cm]
\colhead{} & \colhead{($\mu$Jy)} & \colhead{Limit ($\mu$Jy)} & \colhead{(arcmin$^2$)}}
\decimalcolnumbers
\startdata
F560W & 0.20 & 0.20 & 34.8 \\
F770W (JADES) & 0.021 & 0.036 & 10.1 \\
F770W & 0.19 & 0.18 & 34.5 \\
F1000W & 0.38 & 0.36 & 34.3 \\
F1280W & 0.59 & 0.54 & 34.5 \\
F1500W & 0.68 & 0.63 & 34.5 \\
F1800W & 1.7 & 1.6 & 34.5 \\
F2100W & 2.8 & 2.2 & 34.5 \\
F2550W & 16 & 15 & 34.5  \\
\enddata
\end{deluxetable}

\subsection{Contamination by Spurious Sources}

We expect some fraction of our detected sources to be spurious: not true astrophysical sources, but peaks in the noise. The contribution of these ``false positives" to our number counts is certainly insignificant at the bright end of the distribution, but may become important to the total counts at the faint end.

To quantify the fraction of spurious sources our counts contain, we perform a source extraction on negative images \citep[see] []{Papovich2004}. By inverting the images, the flux of all real sources becomes negative, and the only positive sources remaining in the image are random Gaussian peaks from noise. As the noise is random, the number of sources in the negative image should also reflect the number of spurious sources that will be detected in the original images and removed from the counts.

We invert the mosaic in each filter and mask the noisy edges before performing source detection (using Photutils' find$\_$peaks algorithm), extracting photometry with a $1.2\arcsec$-diameter aperture. We find that the vast majority of spurious sources have integrated fluxes well below our completeness limits, with only a few noise peaks bright enough to contribute to the faint end of the counts. Normalizing by the areas of the SMILES and JADES surveys, we find that we expect $\sim$0.05 to 0.25 counts/arcmin$^2$ from spurious sources at the 80$\%$ completeness limits of SMILES, and even fewer in the JADES parallel. Contributions from spurious sources at brighter fluxes are, as expected, completely negligible, and we therefore make no corrections accounting for their presence.

\section{Results} \label{sec:results}

\subsection{Constructing Number Counts}\label{sec:numbercounts}

Differential and cumulative number counts are derived from the photometric catalogs described in Section~\ref{sec:phot}.
Precise individual areas are measured for each mosaic by counting unmasked pixels (Table~\ref{tab:complete}).  The exposure times are largely uniform across the mosaics: $70-80\%$ of pixels have exposure times within $5\%$ of the median.

We consider counts in each filter down to the $80\%$ completeness limits, which are comparable to the $5\sigma$ point source sensitivities (Table~\ref{tab:complete}). The number counts discussed in the following sections are completeness-corrected, and the error bars reflect purely statistical error.

\subsection{Number Counts at 7.7$\mu$m}

\begin{figure*}
    \centering
    \includegraphics[width=\textwidth]{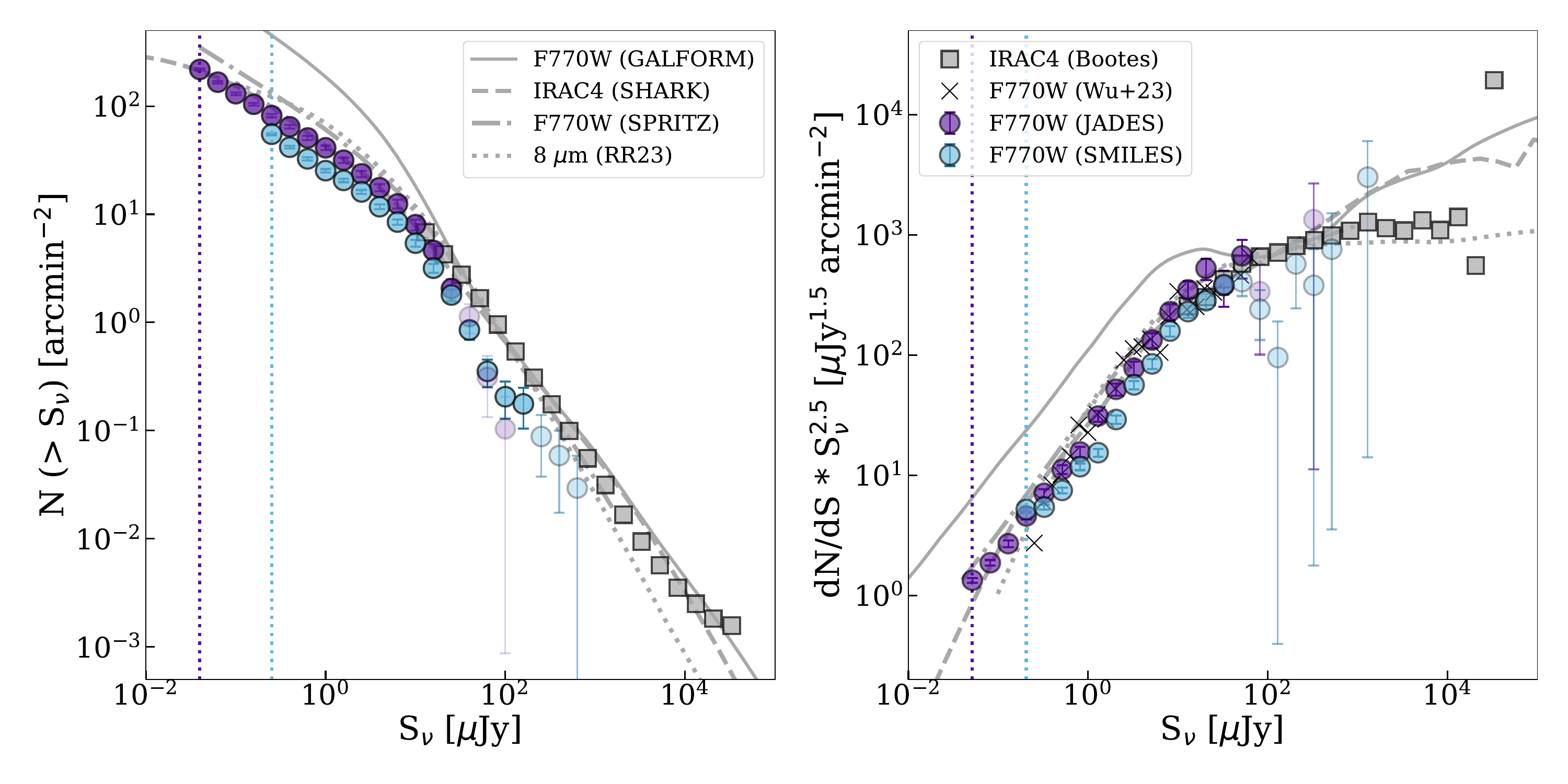}
    \caption{Cumulative (left) and Euclidean-normalized differential (right) number counts at 7.7 (MIRI) and 8 (IRAC) $\mu$m. The deep MIRI F770W counts from the JADES and SMILES surveys are shown as purple and blue circles, respectively, and their 80\% completeness limits are marked by colored, dotted vertical lines. For visual clarity, points at the bright end with high scatter and large error bars are plotted with greater transparency. We also plot the differential F770W counts in two MIRI pointings in the CEERS field measured by \cite{Wu2023} (black crosses in right panel), which agree very well with our data. We link our MIRI data to the bright end by including the IRAC 8 $\mu$m counts from the Spitzer Deep Wide-Field Survey \citep[SDWFS;][]{Ashby2009} as gray squares: together these produce number counts spanning more than five orders of magnitude in flux. The \cite{Ashby2009} IRAC counts have had the contribution from Galactic stars subtracted according to predictions from \cite{Fazio2004}. We also include for comparison predicted number counts at 8 $\mu$m from the GALFORM semi-analytical model \citep[][gray solid line]{Cowley2018}, the SHARK semi-analytical model \citep[][gray dashed line]{Lagos2019}, the SPRITZ simulations \citep[][gray dot-dashed line]{Bisigello2021}, and the models of \cite{RowanRobinson2009, RowanRobinson2023} (gray dotted line). For ease of comparison, we normalize each model to the observed counts at 10$^2$ $\mu$Jy.}
    \label{fig:770counts}
\end{figure*}

Our deepest data is at 7.7 $\mu$m. At low redshift, F770W traces the mid-infrared PAH features, and probes the rest-frame peak of stellar emission at higher redshift, $z \sim 2-5$. At all redshifts, 7.7 $\mu$m---and other mid-infrared---counts will additionally contain a population of dust-obscured AGN and composite (star-forming + AGN) galaxies. When corrected for the contribution of AGN, 7.7 $\mu$m counts can therefore test the distribution of stellar mass, particularly at high redshift, of galaxy formation models. 

The similarity between the MIRI F770W and Spitzer/IRAC Channel 4 (8 $\mu$m) filters also allows us to connect our very deep counts with those from wide-field Spitzer surveys, probing both the bright and faint ends of the distribution. However, the two filters are not identical. While the MIRI F770W filter has slightly higher throughput, the IRAC-4 bandpass is wider on the red side, extending to $\sim9.5$ $\mu$m compared to F770W's $\sim8.75$ $\mu$m. The largest difference between an individual source's IRAC-4 and F770W fluxes will likely occur for sources at redshifts where a bright PAH feature is shifted out of the F770W bandpass but remains in the IRAC-4 bandpass, but this occurs in only two small redshift ranges ($z \sim 0.13-0.27$ and $0.41-0.52$ for the 7.7 and 6.2 $\mu$m PAH features respectively). Early JWST results comparing MIRI to IRAC counts indicate that the effect of this difference is very small \citep[see e.g.][]{Yang2023, Wu2023} in the flux regime where MIRI and IRAC counts overlap.

We present our differential and cumulative 7.7 $\mu$m counts from SMILES and JADES in Figure \ref{fig:770counts}, alongside bright-end IRAC-4 data from the Spitzer Deep Wide-Field Survey (SDWFS) \citep{Ashby2009}. Together, these deep MIRI data and wide Spitzer data provide the deepest 8 $\mu$m counts ever observed, spanning more than five orders of magnitude in flux from $<0.1$ to more than $10^4$ $\mu$Jy. 

The MIRI F770W counts in the SMILES and JADES fields are slightly offset from each other. We have verified that this offset is not due to any differences in the data reduction process. Instead, this is likely due to cosmic variance: the Hubble Ultra Deep Field (HUDF, which forms the center of the SMILES footprint) is known to be slightly underdense, while a number of overdensities have already been catalogued in the small JADES parallel field \citep{Alberts2023}. The difference between the SMILES and JADES counts likely encodes the typical variation that will be observed between surveys \citep[see also the variation in counts for fluxes probed by Spitzer,][]{Bethermin2010}.

The differential counts (right panel of Figure \ref{fig:770counts}), which have been normalized to a Euclidean universe (multiplied by $S_\nu^{2.5}$), display the characteristic power-law decline toward faint fluxes observed by other telescopes, and flatten at the bright end when we remove the contribution from Galactic stars from the model of \cite{Fazio2004}. 

\cite{Wu2023} presented number counts derived from two MIRI pointings in the CEERS field, which we include in Figure \ref{fig:770counts}. They reach an 80\% completeness limit of 0.25 $\mu$Jy in F770W, approximately 1.3 times shallower than our SMILES data and 6.7 times shallower than the MIRI JADES parallel.

For comparison, we also plot in Figure \ref{fig:770counts} the 8 $\mu$m number count predictions from a number of simulations and semi-analytical models of galaxy formation, namely the GALFORM semi-analytical model and GRASIL radiative transfer code \citep[F770W,][]{Cowley2018}, the SHARK semi-analytical model \citep[IRAC-4,][]{Lagos2019}, the SPRITZ simulations \citep[F770W,][]{Bisigello2021}, and the IRAC-4 model of \cite{RowanRobinson2009, RowanRobinson2023}. We discuss these models in greater depth in Section~\ref{models}.

\subsection{Number Counts at 21$\mu$m and $25\mu$m}

\begin{figure*}
    \centering
    \includegraphics[width=\textwidth]{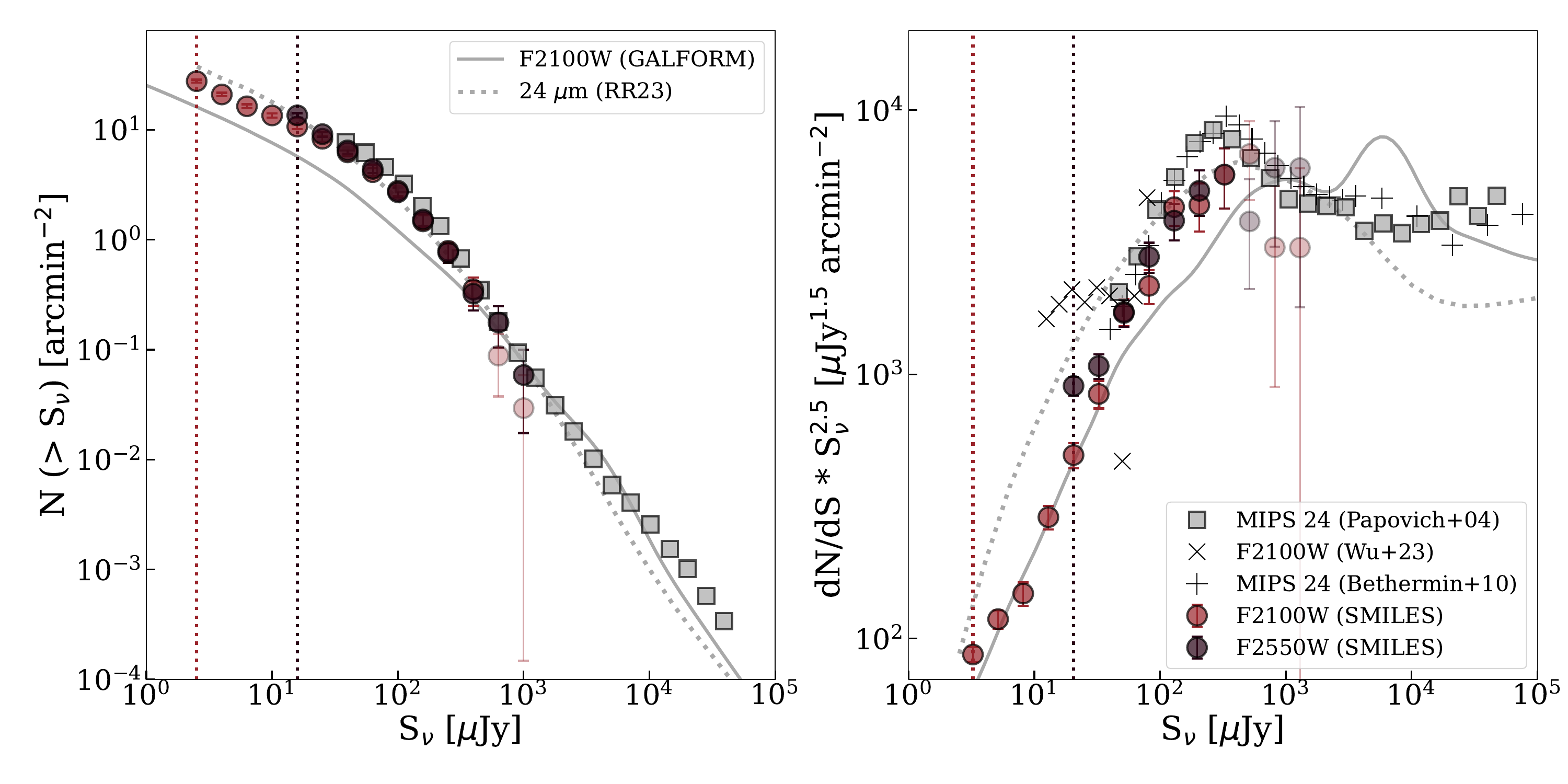}
    \caption{Number counts at 21 and 25 $\mu$m. We plot our cumulative (left) and Euclidean-normalized differential (right) MIRI SMILES F2100W and F2550W counts as colored circles (lighter red and darker brown, respectively). As in Figure \ref{fig:770counts}, points at the bright end with high scatter and large error bars are plotted with greater transparency for visual clarity. Colored, dotted vertical lines mark the 80\% completeness limits of the SMILES 21 and 25 $\mu$m counts. We link to the bright end by including Spitzer/MIPS 24 $\mu$m counts from \cite{Papovich2004} (gray squares). These counts are not significantly contaminated by stars. For comparison, we also plot the differential F2100W counts in a portion of the CEERS field from \cite{Wu2023} at right as black crosses. We additionally include modeled number counts from the literature, both from the GALFORM semi-analytical model \citep[][gray solid line]{Cowley2018} and the models of \cite{RowanRobinson2009, RowanRobinson2023} (gray dotted line). These models have been normalized to the counts at 10$^3$ $\mu$Jy.}
    \label{fig:2100counts}
\end{figure*}

We plot our SMILES cumulative and differential counts at both 21 and 25 $\mu$m in Figure \ref{fig:2100counts}, as the counts in both bands have very similar distributions. Our new JWST counts also align with counts from deep 24 $\mu$m Spitzer surveys \citep[][plotted as gray squares and black plus-signs respectively]{Papovich2004,Bethermin2010}, which we use to extend our distribution to the bright end. Together, these 21/24 $\mu$m counts extend over approximately four orders of magnitude, from less than $100$ to more than $10^5$ $\mu$Jy, and probe deeper than any $\sim$24 $\mu$m counts previously published. Additionally, due to the unprecedented spatial resolution of JWST, our MIRI $\sim$24 $\mu$m counts are not confusion-limited, unlike the Spitzer counts also shown in Figure \ref{fig:2100counts}.

The flux at $\sim25$ $\mu$m probes the mid-infrared warm dust continuum until $z\sim1.5$, when the bright 7.7 $\mu$m PAH feature begins to shift into the band. We observe the characteristic bump at $\sim100$ $\mu$Jy in the Euclidean-normalized differential counts. Despite the minimal overlap of the MIRI F2100W and F2550W bands, the counts display very similar distributions down to the limit of the F2550W data.

The MIPS 24 $\mu$m filter is much broader than the MIRI filters, encompassing most of the wavelength range covered by the MIRI F2100W and F2550W filters. This discrepancy, however, does not appear to severely affect the magnitude or shape of the source counts' distribution, as the MIPS 24 $\mu$m data from \cite{Papovich2004} and \cite{Bethermin2010} display fair agreement, within the uncertainties and scatter, with the MIRI data. No correction has been applied to either the MIRI or MIPS counts to account for the very different bandpasses and bring the counts to alignment.

In addition to the Spitzer counts, we can compare our results to the counts obtained in two MIRI pointings of the CEERS survey by \cite{Wu2023}. While the \cite{Wu2023} counts agree well with ours down to $\sim100$ $\mu$Jy, their differential counts flatten at the faintest end of their data, between $\sim10$ and 100 $\mu$Jy, rather than continuing to decrease as a power-law like our SMILES counts or model predictions. This may be an artifact of those data's lack of background subtraction, as we observe a similar effect to a lesser extent in other bands longward of 10 $\mu$m (discussed below in Section \ref{subsec:otherbands}). \cite{Wu2023} report an F2100W 80\% completeness limit of 13-16 $\mu$Jy in their two pointings, approximately 6-7 times shallower than our SMILES F2100W data.

In general, when normalized near 10$^3$ $\mu$Jy, the models of \cite{RowanRobinson2009, RowanRobinson2023} reasonably closely follow our observed counts down to the faint end, and reproduce the shape of the differential number counts distribution well for fluxes $>$ 100 $\mu$Jy (right panel of Figure \ref{fig:2100counts}). The GALFORM model, on the other hand, underpredicts the number of faint sources in the cumulative counts and does not exhibit the peak in the differential counts near 100 $\mu$Jy seen in both JWST and Spitzer data. Further discussion of these comparisons can be found in Section~\ref{models}.

\subsection{Number Counts in other Bands}\label{subsec:otherbands}

\begin{figure*}
    \centering
    \includegraphics[width=\textwidth]{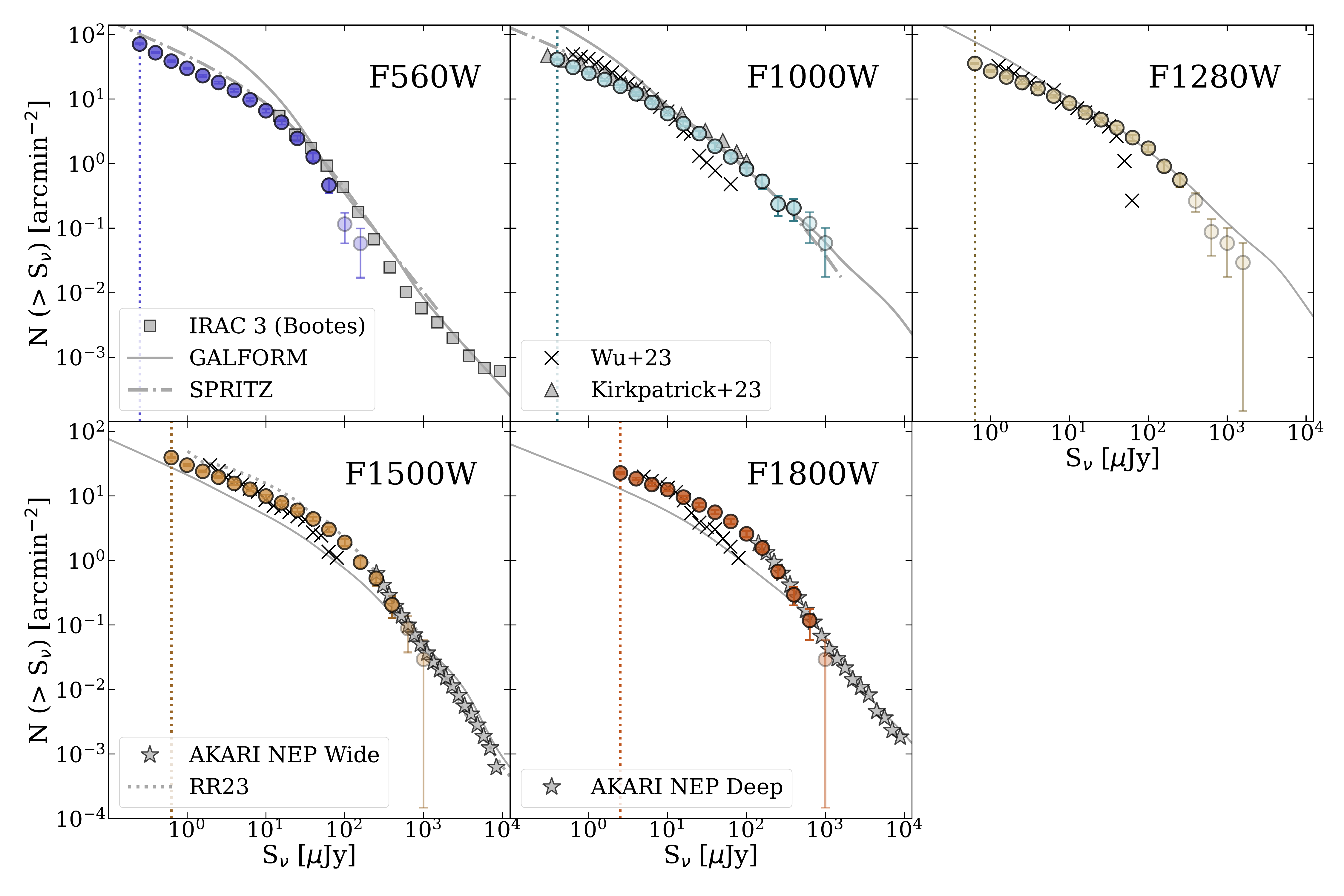}
    \caption{Cumulative number counts for the remaining MIRI bands. MIRI data from the SMILES survey is plotted as colored circles, with colored dotted vertical lines marking the 80\% completeness limit in each band. We also include model number counts in each band from the GALFORM semi-analytical model \citep{Cowley2018} as gray solid lines. All models have been normalized to the counts near the bright end (10$^2$ $\mu$Jy at 5.6, 10, and 12.8 $\mu$m; 10$^3$ $\mu$Jy at 15 and 18 $\mu$m). At 5.6 and 10 $\mu$m, we include the predictions of the SPRITZ simulations \citep[][gray dot-dashed line]{Bisigello2021}. We extend the counts to slightly brighter fluxes by including data from the Spitzer Deep Wide-Field Survey at 5.6 $\mu$m \citep[][gray squares]{Ashby2009} and the AKARI NEP-Wide \citep{Pearson2010} and NEP-Deep surveys \citep{Pearson2014} at 15 and 18 $\mu$m respectively (gray stars). We have corrected the IRAC 5.6 $\mu$m counts for the contribution from Galactic stars \citep{Fazio2004}. At 10, 12.8, 15, and 18 $\mu$m, we include MIRI number counts in two MIRI pointings of the CEERS field as black crosses \citep{Wu2023}. At 10 $\mu$m, counts from more extensive data in the CEERS survey are shown as gray triangles \citep{Kirkpatrick2023}.}
    \label{fig:othercounts}
\end{figure*}

We plot the cumulative number counts for all remaining MIRI bands (F560W, F1000W, F1280W, F1500W, and F1800W) in Figure \ref{fig:othercounts}. We include, when available, other data and models in the bands, including AKARI counts \citep{Pearson2010, Pearson2014}, MIRI counts in the CEERS field \citep{Wu2023, Kirkpatrick2023}, and the GALFORM \citep{Cowley2018}, SPRITZ \citep{Bisigello2021}, and \cite{RowanRobinson2009, RowanRobinson2023} models.

Counts from the AKARI North Ecliptic Pole survey at 15 and 18 $\mu$m \citep{Pearson2010, Pearson2014} extend our MIRI counts to the bright end, and are in very good agreement with our counts where they overlap. As a result, the counts at 15 and 18 $\mu$m extend over nearly four orders of magnitude in flux. These AKARI counts have been corrected for the contributions of stars, but not for the difference between the MIRI F1500W/F1800W and AKARI IRC $L15$/$L18$ bandpasses. The AKARI bandpasses are significantly broader than the corresponding MIRI bandpasses: for example, while the two filters have roughly the same central wavelength, $L15$'s half-power width (6.8 $\mu$m) is more than twice that of F1500W ($\sim3.1$ $\mu$m). Nonetheless, the excellent agreement of the AKARI counts with our MIRI data demonstrates that the different filter shapes do not strongly affect the counts.

At the fainter end, we can compare our counts to the slightly shallower counts measured in two of the MIRI pointings of the CEERS field by \cite{Wu2023}. These counts, in F1000W, F1280W, F1500W, and F1800W, drop off quickly at $\sim 100$ mJy due to the smaller area of the survey, and do not go quite as deep. While they agree fairly well with our counts, some of the bands (particularly F1000W and F1800W) display steeper slopes between 1 and 100 $\mu$Jy. This may be a background subtraction artifact, similar to that observed near the faint end at 21 $\mu$m (right panel of Figure \ref{fig:2100counts}). This hypothesis is reinforced by the 10 $\mu$m counts from \cite{Kirkpatrick2023}: these data, also from the CEERS survey, agree well with our 10 $\mu$m counts.

\subsection{Cosmic Variance}

The error bars in Figures \ref{fig:770counts}, \ref{fig:2100counts}, and \ref{fig:othercounts} reflect only statistical errors. Potentially larger errors occur because of cosmic variance: the differences in source characteristics among different fields on the sky \citep[e.g.,][]{Trenti2008}. Cosmic variance is reduced significantly as the number of distinct fields surveyed increases, and also with their area. In addition, the effects of cosmic variance are more pronounced for the brighter sources than the fainter ones \citep[e.g.,][]{Robertson2010, Driver2010,Moster2011}. 

By combining counts from the large-area, moderate-depth surveys of Spitzer and AKARI with the (to date) small areas surveyed to great depth by JWST, the cosmic variance effects at play can be understood and mitigated. We discuss cosmic variance for each wavelength in turn.

{\bf 5.6 $\mu$m}:
The top left panel of Figure \ref{fig:othercounts} compares our 5.6 $\mu$m counts to those in IRAC-3 reported in $\sim$ eight deg$^2$ in the Bootes field by \cite{Ashby2009}. The \cite{Ashby2009} counts demonstrate excellent agreement with those in the 0.38 deg$^2$ survey in the Extended Groth Strip surveyed to $\sim$ five times deeper reported by \cite{Barmby2008}, and with counts from the IRAC Shallow Survey \citep{Fazio2004}. The large areas surveyed and the agreement of counts from two widely separated fields indicate that cosmic variance does not have a strong effect on these counts. Their close agreement with our JWST-based counts in the region of overlap indicates, despite the small area surveyed, that the latter counts are also representative without a large bias due to cosmic variance.

{\bf 7.7 $\mu$m}:
Figure~\ref{fig:770counts} compares our counts at 7.7 $\mu$m to those from \cite{Ashby2009} at 7.8 $\mu$m. As at 5.6 $\mu$m, the \cite{Ashby2009} 7.8 $\mu$m counts agree well with counts measured in other fields with Spitzer \citep{Barmby2008, Fazio2004}, and agree with our MIRI F770W counts where they overlap. We therefore do not expect cosmic variance to seriously bias our 7.7 $\mu$m counts.

{\bf 15 \& 18 $\mu$m}:
The counts in Figure~\ref{fig:othercounts} at 15 and 18 $\mu$m use results from the AKARI North Ecliptic Pole (NEP) surveys\footnote{ISO counts at 15 $\mu$m were reported by \cite{Oliver1997} but have been supplanted by those from AKARI.}. These observations were designed to reduce cosmic variance to the $5 - 10$\% level \citep{Pearson2010}\footnote{The coverage of even larger (and more) fields in IRAC Band 4 and MIPS 24 $\mu$m reinforces our expectation that these surveys are not significantly affected by cosmic variance.}. They cover areas of $\sim$ 0.4 - 0.6 deg$^2$ and 5.8 deg$^2$ to depths of 117 (150) and 250 (300) $\mu$Jy respectively at 15 (18) $\mu$m \citep{Pearson2010, Pearson2014}. The excellent agreement with the counts in the SMILES field in the region of overlap again implies that the entire range shown in Figure~\ref{fig:othercounts} is unlikely to be compromised by cosmic variance.  On the other hand, MIRI counts from \cite{Wu2023} fall low at the bright end, suggestive of an influence of variance on those results due possibly to the significantly smaller field available ($\sim$ 6 arcmin$^2$). 

{\bf 21 - 25 $\mu$m}:
Figure~\ref{fig:2100counts} compares JWST/MIRI counts at 21 and 25.5 $\mu$m with Spitzer/MIPS counts at 24 $\mu$m. The counts from \cite{Papovich2004} are from five widely separated fields with $\sim$ 0.61 deg$^2$ surveyed to a depth of $\sim$ 80 $\mu$Jy and $>$ 10 deg$^2$ to a depth of at least 270 $\mu$Jy. \cite{Bethermin2010} report surveys of $\sim$ 0.64 deg$^2$ to $\sim$ 80 $\mu$Jy and 53 deg$^2$ to at least $\sim$ 400 $\mu$Jy in seven additional fields. These surveys agree very well with each other and with our counts down to their limits of 80 $\mu$Jy (see Figure~\ref{fig:2100counts}). Again, this indicates that the full range of counts presented here are not strongly affected by cosmic variance. 

\begin{deluxetable*}{c|ccccccccc}
\tablecaption{Cumulative MIRI number counts from SMILES and JADES \label{tab:counts}}
\tablewidth{0pt}
\tablehead{
\colhead{Flux} & \colhead{} & \colhead{F770W} & \colhead{F770W} & \colhead{} & \colhead{} & \colhead{} & \colhead{} & \colhead{} & \colhead{}\\[-0.25cm]
\colhead{($\mu$Jy)} & \colhead{F560W} & \colhead{(SMILES)} & \colhead{(JADES)} & \colhead{F1000W} & \colhead{F1280W} & \colhead{F1500W} & \colhead{F1800W} & \colhead{F2100W} & \colhead{F2550W}}
\decimalcolnumbers
\startdata
3.98e-2 & - & - & 218 $\pm$ 5 & - & - & - & - & - & - \\
6.31e-2 & - & - & 167 $\pm$ 4 & - & - & - & - & - & - \\
1.00e-1 & - & - & 130 $\pm$ 4 & - & - & - & - & - & - \\
1.58e-1 & - & - & 104 $\pm$ 3 & - & - & - & - & - & - \\
2.51e-1 & 71.3 $\pm$ 1.4 & 55.0 $\pm$ 1.3 & 81.8 $\pm$ 2.9 & - & - & - & - & - & - \\
3.98e-1 & 52.2 $\pm$ 1.2 & 41.7 $\pm$ 1.1 & 64.5 $\pm$ 2.6 & 41.3 $\pm$ 1.1 & - & - & - & - & - \\
6.31e-1 & 38.7 $\pm$ 1.1 & 32.6 $\pm$ 1.0 & 50.8 $\pm$ 2.3 & 31.1 $\pm$ 1.0 & 35.6 $\pm$ 1.0 & 39.3 $\pm$ 1.0 & - & - & - \\
1.00e+0 & 30.0 $\pm$ 0.9 & 25.3 $\pm$ 0.9 & 41.2 $\pm$ 2.0 & 25.0 $\pm$ 0.9 & 27.1 $\pm$ 0.9 & 30.0 $\pm$ 0.9 & - & - & - \\
1.58e+0 & 23.0 $\pm$ 0.8 & 20.6 $\pm$ 0.8 & 31.6 $\pm$ 1.8 & 20.0 $\pm$ 0.8 & 22.0 $\pm$ 0.8 & 24.1 $\pm$ 0.8 & - & - & - \\
2.51e+0 & 18.0 $\pm$ 0.7 & 16.1 $\pm$ 0.7 & 23.6 $\pm$ 1.6 & 15.8 $\pm$ 0.7 & 18.0 $\pm$ 0.7 & 19.6 $\pm$ 0.8 & 22.7 $\pm$ 0.8 & 27.7 $\pm$ 0.9 & - \\
3.98e+0 & 13.7 $\pm$ 0.6 & 11.7 $\pm$ 0.6 & 17.6 $\pm$ 1.3 & 12.1 $\pm$ 0.6 & 14.5 $\pm$ 0.6 & 15.6 $\pm$ 0.7 & 18.5 $\pm$ 0.7 & 21.0 $\pm$ 0.8 & - \\
6.31e+0 & 9.73 $\pm$ 0.53 & 8.48 $\pm$ 0.50 & 12.5 $\pm$ 1.1 & 8.81 $\pm$ 0.51 & 11.2 $\pm$ 0.6 & 12.7 $\pm$ 0.6 & 15.0 $\pm$ 0.7 & 16.4 $\pm$ 0.7 & - \\
1.00e+1 & 6.60 $\pm$ 0.44 & 5.41 $\pm$ 0.40 & 8.01 $\pm$ 0.90 & 5.96 $\pm$ 0.42 & 8.67 $\pm$ 0.50 & 9.96 $\pm$ 0.54 & 12.6 $\pm$ 0.6 & 13.5 $\pm$ 0.6 & - \\
1.58e+1 & 4.37 $\pm$ 0.35 & 3.17 $\pm$ 0.30 & 4.62 $\pm$ 0.69 & 4.16 $\pm$ 0.35 & 6.17 $\pm$ 0.42 & 7.81 $\pm$ 0.48 & 9.58 $\pm$ 0.53 & 10.7 $\pm$ 0.6 & 13.5 $\pm$ 0.6\\
2.51e+1 & 2.46 $\pm$ 0.27 & 1.79 $\pm$ 0.23 & 2.06 $\pm$ 0.46 & 2.92 $\pm$ 0.29 & 4.79 $\pm$ 0.37 & 5.99 $\pm$ 0.42 & 7.26 $\pm$ 0.46 & 8.29 $\pm$ 0.49 & 9.13 $\pm$ 0.51\\
3.98e+1 & 1.27 $\pm$ 0.19 & 0.85 $\pm$ 0.16 & 1.13 $\pm$ 0.34 & 1.86 $\pm$ 0.23 & 3.58 $\pm$ 0.32 & 4.41 $\pm$ 0.36 & 5.58 $\pm$ 0.40 & 6.23 $\pm$ 0.43 & 6.50 $\pm$ 0.43\\
6.31e+1 & 0.46 $\pm$ 0.12 & 0.35 $\pm$ 0.10 & 0.31 $\pm$ 0.18 & 1.27 $\pm$ 0.19 & 2.53 $\pm$ 0.27 & 3.03 $\pm$ 0.30 & 4.03 $\pm$ 0.34 & 4.12 $\pm$ 0.35 & 4.41 $\pm$ 0.36\\
1.00e+2 & 0.12 $\pm$ 0.06 & 0.21 $\pm$ 0.08 & 0.10 $\pm$ 0.10 & 0.83 $\pm$ 0.16 & 1.73 $\pm$ 0.22 & 1.91 $\pm$ 0.24 & 2.58 $\pm$ 0.27 & 2.79 $\pm$ 0.28 & 2.70 $\pm$ 0.28\\
1.58e+2 & 0.06 $\pm$ 0.04 & 0.18 $\pm$ 0.07 & - & 0.53 $\pm$ 0.12 & 0.91 $\pm$ 0.16 & 0.94 $\pm$ 0.17 & 1.56 $\pm$ 0.21 & 1.47 $\pm$ 0.21 & 1.53 $\pm$ 0.21\\
2.51e+2 & - & 0.09 $\pm$ 0.05 & - & 0.24 $\pm$ 0.08 & 0.56 $\pm$ 0.13 & 0.53 $\pm$ 0.12 & 0.68 $\pm$ 0.14 & 0.79 $\pm$ 0.15 & 0.76 $\pm$ 0.15 \\
3.98e+2 & - & 0.06 $\pm$ 0.04 & - & 0.21 $\pm$ 0.08 & 0.26 $\pm$ 0.09 & 0.21 $\pm$ 0.08 & 0.29 $\pm$ 0.09 & 0.35 $\pm$ 0.10 & 0.32 $\pm$ 0.10 \\
6.31e+2 & - & 0.03 $\pm$ 0.03 & - & 0.12 $\pm$ 0.06 & 0.09 $\pm$ 0.05 & 0.09 $\pm$ 0.05 & 0.12 $\pm$ 0.06 & 0.09 $\pm$ 0.05 & 0.18 $\pm$ 0.07 \\
1.00e+3 & - & - & - & 0.06 $\pm$ 0.04 & 0.06 $\pm$ 0.04 & 0.03 $\pm$ 0.03 & 0.03 $\pm$ 0.03 & 0.03 $\pm$ 0.03 & 0.06 $\pm$ 0.04 \\
\enddata
\tablecomments{All reported counts (2-10) are cumulative (N $>$ $S_{\nu}$), with units arcmin$^{-2}$.}
\end{deluxetable*}

\section{Discussion} \label{sec:discussion}

\subsection{Models}
\label{models}
Figures 2 - 4 show the predictions of number count models between 5 and 25 $\mu$m, many of which diverge from observed counts. Many models, even those published recently, use SED templates created before Spitzer IRS data were available and/or use relatively abstract templates that do not match the results  with full Spitzer data \citep[e.g., as in][]{Rieke2008, Magdis2012, Schreiber2018, Bernhard2021}. This could be important because the aromatic bands dominate the infrared output of star-forming galaxies from 6 through 12 $\mu$m and fall within the wavelength range of JWST and Spitzer number counts for $z\sim0-3$. For example, \citet{Cowley2018} utilize a combination of models from GRASIL \citep{Silva1998} and GALFORM \citep{Granato2000}. However, in these models the PAH bands are represented very schematically, and in the latter one the  region between 15 and 40 $\mu$m is not represented accurately and the silicate absorption features at 10 and 18 $\mu$m are not included. The modeled number counts by \cite{RowanRobinson2023} are based on mid-infrared SEDs from \cite{Rowanrobinson2004, Rowanrobinson2005}. These are constrained by Spitzer photometry, but the PAH bands are from ISO spectroscopy of a very small number of galaxies \citep{Rowanrobinson2001} and again are quite schematic. This relatively limited sample of PAH spectra can cause problems: for an example, compare the proposed spectra in Figure 6 in \citet{Rowanrobinson2005} for ``Arp 220 starbursts" with the actual spectrum of Arp 220 from \cite{Spoon2004}. This approach to modeling may be responsible for the deviation between the \cite{RowanRobinson2023} model and the observed counts distribution at the faint end of Figure \ref{fig:2100counts}. Unfortunately, predicted counts from this model are not available at wavelengths between 8 and 24 $\mu$m.

Two sets of models, SHARK \citep{Lagos2019} and SPRITZ \citep{Bisigello2021}, do include accurate spectra of the PAH bands making use of Spitzer IRS spectra. The SHARK models are based on a semi-analytic model of galaxy formation and evolution. These models are compared with counts in the IRAC bands, and their good matching of the 7.7 $\mu$m counts presented here to very faint levels is very encouraging in terms of such models. Obviously, using this approach to model counts at the longer wavelengths is very desirable, continuing with accurate template SEDs.

The SPRITZ models are empirically based on a global analysis of galaxy properties. They match the observed counts very well at 5.6, 7.7, and 10 $\mu$m, including to our very deep 7.7 $\mu$m counts. For both sets of models, the agreement with our counts at shorter wavelengths indicates that representing the PAH bands accurately is critical to interpreting the number counts presented in this paper; this could be confirmed with predicted model counts in the longer MIRI bands, which are currently not available. 

\subsection{The Cosmic Infrared Background}

The Cosmic Infrared Background (CIB) is the fraction of the Extragalactic Background Light (EBL) emitted in the infrared, and arises mainly from dust-obscured star formation (with contributions from dust-obscured AGN) in galaxies across cosmic time \citep{Hauser1998, Gispert2000, Lagache2005}. With more sensitive telescopes, more and more of the CIB can be resolved into individual sources, e.g. the sources making up our galaxy number counts. Our discussion below follows on from the treatment in \citet{Driver2016}.

To estimate the total 8 $\mu$m flux from sources detectable in our MIRI data, we combine and integrate 8 $\mu$m number counts from the Spitzer Deep Wide-Field Survey \citep[SDWFS,][]{Ashby2009} at the bright end, with stars removed, and our JADES MIRI parallel at the faint end. To combine the data sets, we exclude all JADES MIRI parallel counts at fluxes brighter than $S_{\nu} = 50$ $\mu$Jy, because the counts in this flux regime are affected by cosmic variance due the small area of the survey. The wider-area SDWFS has data in this flux range, which will be less affected by cosmic variance. Correspondingly, we exclude all counts at fluxes fainter than $S_{\nu} = 30$ $\mu$Jy from the SDWFS IRAC data: this flux corresponds to the 8 $\mu$m SDWFS 80\% completeness limit. In the small remaining region of overlap, the MIRI and IRAC counts agree well (see Figure \ref{fig:770counts}). We also do not include the two brightest SDWFS IRAC points ($S_{\nu} > 2\times10^4$ $\mu$Jy) in our integration, as they deviate sharply from the flat, Euclidean trend expected at very bright fluxes in the right panel of Figure \ref{fig:770counts} due to cosmic variance. After integrating and propagating through the statistical errors on each bin, we obtain an 8 $\mu$m CIB flux of $3.02\pm0.09$ nW m$^{-2}$ sr$^{-1}$, slightly larger than estimates from 8 $\mu$m Spitzer surveys \citep[2.6 nW m$^{-2}$ sr$^{-1}$,][]{Fazio2004}.

\citet{Driver2016} estimated the CIB at 8 $\mu$m using Spitzer/IRAC data from \citet{Barmby2008}, who surveyed 0.38 deg$^{2}$ in the Extended Groth Strip. Their uncertainty from cosmic variance ($\pm 0.08$ mag)  translates to approximately $\pm 0.035$ mag for the 10 deg$^2$ area we have used to anchor our counts at bright levels \citep{Ashby2009}. They overstate the zero point error; the absolute calibration of the IRAC data is accurate to $\sim$ 3\%. The ultra-deep JWST data basically remove uncertainties due to extrapolation to low fluxes. The net error in our value should therefore be $\sim$ 10\%.

We calculate number counts at 21/24 $\mu$m in the same manner, by combining Spitzer/MIPS counts \citep{Papovich2004} and our new MIRI SMILES counts. We remove the brightest MIRI 21 $\mu$m points ($S_{\nu} > 200$ $\mu$Jy) which are most affected by cosmic variance. The 80\% completeness limit for the \cite{Papovich2004} MIPS data lies at $S_{\nu}\sim100$ $\mu$Jy, so we do not include any fainter MIPS counts in our integration. The SMILES/MIRI and MIPS points agree very well (see Figure \ref{fig:2100counts}) near this remaining small region of overlap. We integrate from the faintest MIRI point (3.25 $\mu$Jy) to the brightest MIPS point (4.76$\times10^4$ $\mu$Jy), and do not extrapolate; this integrated value is therefore a lower limit on the 21/24 $\mu$m CIB flux. Propagating through only the statistical errors on the counts, we obtain a flux of $2.85 \pm 0.04$ nW m$^{-2}$ sr$^{-1}$. Since we have taken our bright 24 $\mu$m counts from the same source as \citet{Driver2016}, their estimate of a cosmic variance error of $\pm 0.07$ also applies to our result. Most other errors listed in \citet{Driver2016} should be reduced by use of the JWST data. As before, the zero point error is significantly overstated; the absolute calibration of MIPS is accurate to $\sim$ 3\% \citep{Rieke2008}. We  conclude that the total error in our value is, as at 7.8 $\mu$m, $\sim$10\%.

\cite{Bethermin2010} found a CIB flux of 2.29$\pm$0.09 nW m$^{-2}$ sr$^{-1}$ from Spitzer 24 $\mu$m counts probing fluxes $>35 \mu$Jy. However, when the counts were extrapolated at the faint end, they calculated an integrated 24$\mu$m CIB flux of 2.86$^{+0.19}_{-0.16}$ nW m$^{-2}$ sr$^{-1}$, in excellent agreement with our value. \citet{Dole2006} also measured the 24 $\mu$m CIB by adding up the flux from all sources brighter than 60 $\mu$Jy and using the counts of \citet{Papovich2004} to determine the contribution from fainter sources, finding an EBL value of 2.7 $\pm$ 0.26 nW m$^{-2}$ sr$^{-1}$. Because \cite{Dole2006} do not quote any additional error from the extrapolation to very faint limits, this uncertainty is likely somewhat underestimated. 

TeV gamma rays are attenuated as they traverse intergalactic space through pair production with the ambient ultraviolet, optical, and infrared photons. Semi-analytic models (SAMs) of the history of galaxy formation and galaxy properties can be combined with TeV gamma ray measurements to predict the spectrum of the EBL. Our values at both 7.8 and 24 $\mu$m are in mild tension ($\sim$3$\sigma$) with the fiducial EBL model based on TeV gamma ray sources by \citet{Gilmore2012}, which predicts values of 2.4 $\pm$ 0.2 nW m$^{-2}$ sr$^{-1}$ at both wavelengths. Such a tension has not been previously noted in the literature, because EBL measurements from Spitzer previously available for comparison had much larger uncertainties. On the other hand, our EBL measurements agree with the model of \cite{Desai2019}, which predicts 3.1 $\pm$ 1.0 and 3.1 $\pm$ 0.6 nW m$^{-2}$ sr$^{-1}$ at 7.8 and 24 $\mu$m respectively. As shown by \citet{Abeysekara2019}, the highest weight values based on TeV gamma ray observations tend to fall on or below the Gilmore fiducial, although the disagreement with our values is not huge. It should be kept in mind that the EBL estimates from source counts are strictly lower limits, since they exclude any diffuse emission and discriminate against extended sources. These accurate determinations of the mid-IR EBL suggest that some tuning of the SAMs used to predict the EBL may be desirable.

\section{Summary and Conclusions}\label{sec:conclusions}

We measure JWST/MIRI source counts from 5 to 25 $\mu$m using data from the SMILES and JADES surveys. We combine these unprecedentedly deep counts with existing, wider data from Spitzer and AKARI to produce counts spanning, in some bands, five or more orders of magnitude in flux. The uncertainties on our counts from cosmic variance and spurious sources are small.

When comparing our observed counts to predictions from a number of semi-analytical models of galaxy formation, we find that the accuracy of the predicted counts is highly dependent on the templates used to construct model SEDs. In particular, models incorporating Spitzer/IRS spectroscopy of the PAH features from 6-12 $\mu$m more closely match our observed counts.

We integrate our deep JWST + wider-area Spitzer source counts at 8 and 24 $\mu$m to constrain the cosmic infrared background at these wavelengths. Our results are consistent with those from shallower infrared surveys which extrapolated their data at the faint end, but our deep data significantly reduces the associated uncertainty on the integrated EBL. Our results are strictly lower limits since our number counts are not sensitive to diffuse or extended emission. Nonetheless,  we obtain EBL fluxes larger (by $\sim$3$\sigma$) than some predicted by fiducial EBL models based on TeV blazar data, but consistent with others. This suggests that some tuning of the semi-analytic models used for these predictions may be needed.

Robust infrared source counts spanning several orders of magnitude in flux can also be directly compared to predicted counts from models of galaxy formation. State-of-the-art models have large uncertainties at wavelengths longer than 8 $\mu$m. Our accurate counts should encourage generation of more robust models for comparison. Such modeling is also timely, since JWST's unique capabilities compared to any previous infrared telescope will allow it to continue to provide critical constraints on models of galaxy formation and extragalactic light as larger areas of the sky are imaged to unprecedented depth.

\begin{acknowledgments}
MS, SA, GR, JL, and YZ acknowledge support from the JWST Mid-Infrared Instrument (MIRI) grant 80NSSC18K0555, and the NIRCam science support contract NAS5-02105, both from NASA Goddard Space Flight Center to the University of Arizona. The JWST data presented in this paper were obtained from the Mikulski Archive for Space Telescopes (MAST) at the Space Telescope Science Institute. The observations can be accessed via\dataset[DOI.]{http://dx.doi.org/10.17909/zfxs-ye40}
PGP-G acknowledges support from grant PID2022-139567NB-I00 funded by Spanish Ministerio de Ciencia e Innovaci\'on MCIN/AEI/10.13039/501100011033, FEDER {\it Una manera de hacer Europa}.
AJB acknowledges funding from the ``FirstGalaxies" Advanced Grant from the European Research Council (ERC) under the European Union's Horizon 2020 research and innovation program (Grant agreement No. 789056). We thank the anonymous referee for their helpful comments which improved the paper.
\end{acknowledgments}

\vspace{5mm}
\facilities{JWST(MIRI)}

\software{Astropy \citep{Astropy2013,Astropy2018,Astropy2022}, Matplotlib \citep{Hunter2007}, NumPy \citep{VanDerWalt2011}, photutils \citep{Bradley2022}, WebbPSF \citep{Perrin2014}}

\bibliography{main}{}
\bibliographystyle{aasjournal}

\end{document}